\documentstyle[12pt,aasms4]{article}

\lefthead{Laws and Gonzalez}
\righthead{16 Cygni A and B}
\slugcomment{Accepted for Publication in the Astrophysical Journal}
\begin{document}

\title{A Differential Spectroscopic Analysis of 16 Cygni A and B}
\author{Chris Laws and Guillermo Gonzalez}
\affil{Astronomy Department, University of Washington, P.O. Box 351580, 
Seattle, WA 98195 USA}
\authoremail{laws@astro.washington.edu, gonzalez@astro.washington.edu}

\begin{abstract}

We utilize high-resolution, high signal-to-noise spectra to perform a 
differential analysis of Fe abundances in the common proper-motion pair 16 
Cyg A and B. We confirm that both stars are slightly metal-rich compared to 
the Sun, and we show for the first time that the primary is enhanced in Fe 
relative to the secondary by a significant amount. We find 
$\Delta$[Fe/H]$=+0.025\pm0.009$.  This tends to support the ``self-pollution'' 
scenario proposed by Gonzalez (1998), though lack of a complete 
understanding of small primordial metallicity variations among binaries and 
open cluster members prevents a definitive conclusion.

\end{abstract}


\section{Introduction}

The nearby common proper-motion pair 16 Cyg A and B (HD\,186408 and HD\,186427, 
respectively) have been of particular interest to stellar astronomers for a 
number of reasons. In their physical characteristics, they are both very 
similar to the Sun and as such have been labeled ``solar twins'' or ``solar 
analogs'' (Friel et al. 1993, hereafter F93). Like any human 
twins, however, these two stars clearly cannot be exactly similar to the Sun 
or even to each other, and it is their differences that have generated the 
most recent interest. Most notably, radial-velocity studies have revealed that 
16 Cyg B harbors a planetary-mass companion, while 16 Cyg A apparently does 
not (Cochran et al. 1997). This remarkable observation has itself been invoked by 
some to explain other unexpected chemical differences between these two stars 
(Cochran et al. 1997; King 1997; Gonzalez 1998, hereafter G98) -- their Li 
abundances differ by a factor of five of more (Deliyannis et al. 2000, 
hereafter D00), while studies of the Fe abundance of this pair have 
consistently suggested that 16 Cyg A may be slightly more metal-rich 
($\sim$0.05 dex) than its companion (G98). How are these differences achieved 
in such otherwise similar stars, which share a presumably common primordial 
environment? To what level can dynamic interactions with a planetary system 
affect abundances in a stellar photosphere?

Unfortunately, with the exception of lithium, the small differences in line
depths between these two stars combined with the relatively large uncertainties
in abundance determinations to date have prevented any conclusive statements
to be made about deviations between 16 Cyg A and B with regard to any other 
element, including iron (G98; D00); within the uncertainties, the measured 
chemical differences are all effectively zero. The primary goal of this work 
is to utilize a new differential line abundance analysis method to better 
constrain the relative [Fe/H] values between 16 Cyg A and B. Such a method, 
as shown by Langer et al. (1998), eliminates non-trivial uncertainties in 
oscillator strengths, $gf$ values. Further, we utilize a more extensive Fe 
linelist covering a wider range in lower excitation potential ($\chi_{\rm l}$) 
than previous published studies. In \S 2, we describe our observations and 
the data reduction, in \S 3 we describe our analysis methods and present 
results, and we discuss their significance in \S 4.

\section{Observations}

Our intentions from the start have been to perform a very careful spectroscopic
analysis of 16 Cyg A and B, optimized for differential analysis. We determined
at the outset to make every
effort to minimize any systematic differences which might arise due to variations
in our observations, reductions, and subsequent analyses of each star. With this
goal in mind, we obtained optical spectra of 16 Cyg A and B within 30 minutes of
each other on December 22, 1999, with the McDonald Observatory 2.7m telescope. We 
employed the coud\'{e} echelle spectrograph (Tull et al. 1995) and a 2048x2048 
Tektronix CCD, and made no changes to the instrument between the two 
observations. We achieved a resolving power of 58,000 (as measured on a
Th-Ar lamp spectrum) with a signal-to-noise ratio (S/N) of $\sim$400 per 
pixel at 6700 \AA. Spectra of a hot star were also obtained within one hour
and at a similar airmass in order to compensate for telluric features in the
spectra of 16 Cyg A and B.

Data for 16 Cyg A and B were reduced in as nearly an identical manner as 
possible, in both cases following the general method described in Gonzalez 
(1997). Each image was processed with the same bias and flat fields, and 
corrections for scattered light were made with the same fitting functions in 
both spectra. In addition, continua of individual orders were normalized with 
exactly the same functions in both stars. These steps were undertaken to 
minimize possible systematic differences between line depths which would 
ultimately impact our differential analysis. Our efforts yielded continuous 
one-dimensional spectra in the blue up to 5500 \AA, with gaps thereafter up to 
roughly 10,000 \AA.

\section{Analysis}
\subsection{Equivalent Widths}

To eliminate potential bias in measuring the equivalent widths (EWs) of 
spectral features, one of us (GG) renamed each spectrum, and the other (CL) 
performed the measurements and analysis with no prior knowledge of which 
spectrum was associated with which star. We employed a list of high-quality
atomic lines used in our previous studies of planet bearing stars, and
measured the EWs of 60 Fe I and 8 Fe II lines spanning broad ranges in $\chi_
{\rm l}$ from 0.09 to 6.22 eV and in EW from 4.7 to 119.8 m\AA. In every case
the EW was measured for a line first in one star, then immediately afterwards
in the second star before moving on to the next line. This procedure was
followed to minimize any systematic differences in EW measurements between the
two stars. Telluric contamination in some Fe lines was addressed by dividing
the science object spectra by that of a hot star, and where performed, this
division, as well as considerations of blending by nearby features, was treated
identically for each line in each star. Our linelist and measured EW values are
presented in Table 1.

\subsection{Standard Analysis}

We employed the measured EWs from \S 3.1 and associated $gf$ values taken 
from Gonzalez et al. (2000) to derive a set of basic stellar parameters 
($T_{\rm eff}$, $\log g$, $\xi_{\rm t}$, and [Fe/H]) and associated 
uncertainties for 16 Cyg A and B, using the methods described extensively in 
G98 and Gonzalez \& Vanture (1998). Briefly, all four of these parameters were
systematically iterated with a recent version of the LTE abundance code MOOG
(Sneden, 1973) using Kurucz model atmospheres (1993) until the mean Fe I and Fe II
abundances were equal and correlations of the individual Fe I line abundances
with both $\chi_{\rm l}$ and the logarithm of reduced equivalent width (REW) were zero.

Following the method described in Gonzalez \& Vanture (1998), uncertainties in
$\xi_{\rm t}$ were determined from the standard deviation in the slope of a
least-squares fit to the Fe I vs. $\log (REW)$ data. This resulting uncertainty in $\xi_{
\rm t}$ was summed in quadrature with the standard deviation in the slope of
a least-squares fit to the Fe I vs. $\chi_{\rm l}$ data to estimate the uncertainty
in $T_{\rm eff}$. We combined theses uncertainties in $\chi_{\rm l}$ and $T_{
\rm eff}$ with the observed scatter in modeled Fe I line abundances to calculate
our final uncertainty in [Fe/H].\footnote {Throughout this paper, all stated
uncertainties are at the 1-$\sigma$ level.}

Our results using these standard methods are presented in Table 2, and we
confirm previous studies: 16 Cyg A and B are very similar to each other, as
well as to the Sun. The pair are each slightly metal-enhanced relative to
solar, however, with 16 Cyg A presenting an 0.03 dex overabundance of Fe
relative to 16 Cyg B, although this difference is not significant given our
formal estimates of uncertainty.

\subsection{Differential Analysis}

To reduce the uncertainties in our estimates from \S 3.2, we reanalyzed our
EW data from these two stars employing a technique similar to that described
and effectively utilized by Langer et al. (1998) in searching for small
metallicity variations amongst red giants in M92. Essentially, this method
mirrors the ``standard" method discussed above, but instead of determining
chemical abundances for each star by averaging the results determined from
individual lines and subsequently comparing these averages between stars, it
determines differential chemical abundances between stars by taking the
differences in abundances calculated for each line {\it individually}. This
differential strategy effectively eliminates uncertainties in $gf$ values,
and allows one to take advantage of the increased precision in $\Delta$[Fe/H]
to further constrain differential values of the other stellar parameters.

We used our values for $T_{\rm eff}$, $\log g$, and $\xi_{\rm t}$ from \S 3.2 
to calculate [Fe/H] for each line in each star, then determined differential 
Fe I abundances ($\Delta$[Fe/H]) for each line individually.\footnote{In the 
following discussion, all ``$\Delta$'' values are of the form $\Delta X = 
X_{\rm 16 Cyg A} - X_{\rm 16 Cyg B}$.} If the $\Delta$$T_{\rm eff}$ and 
$\Delta \xi_{\rm t}$ values from \S 3.2 were both correct, then plots of 
$\Delta$[Fe/H] vs. both $\chi_{\rm l}$ and $\log (REW)$ would be uncorrelated. Our
initial such plots using these ``$\Delta$'' values as inputs, however, showed
slopes of $0.003 \pm 0.002$ dex/eV for $\Delta$[Fe/H] vs. $\chi_{\rm l}$ and 
$-0.010 \pm 0.009$ dex/$\log (REW)$ for $\Delta$[Fe/H] vs. $\log (REW)$ (for comparison,
these values in our standard analysis of 16 Cyg A in \S 3.2 were $0.001 \pm
0.006$ and $0.003 \pm 0.024$, respectively).
While these results are only slightly larger than their standard
uncertainties, we note that they are both more precise and appreciably larger
in magnitude than the final ``zero-slope'' conditions which resulted from our
standard analysis. Such correlations indicate that the values of $\Delta$
$T_{\rm eff}$ and $\Delta \xi_{\rm t}$ calculated from the results in Table 2 are
off by some amount, and so we iterated $\Delta$$T_{\rm eff}$ and $\Delta
\xi_{\rm t}$ by changing $T_{\rm eff(16 Cyg B)}$ and $\xi_{\rm t(16 Cyg B)}$
until these correlations were forced to zero.\footnote{Additionally, one consistently 
discrepant Fe I line at 6864.32 \AA\ was discarded.} Figure 1 presents 
a visualization of the effect of changing $\Delta$$T_{\rm eff}$ and $\Delta 
\xi_{\rm t}$ on the $\chi_{\rm l}$ and $\log (REW)$ slopes. For $\log g = 4.21$ 
and 4.26 for 16 Cyg A and B, respectively, we find our best solution at 
$\Delta$$T_{\rm eff} = 62 \pm 14$K and $\Delta \xi_{\rm t} = 0.05 \pm 0.01$ 
km/s, where the stated uncertainties were calculated in precisely the same
manner as described in \S 3.2. We note that these values are in very good 
agreement with our earlier, though less precise, results generated via the 
``standard'' method.

Figures 2 (a) and (b) show our final plots of $\Delta$[Fe/H] vs. both 
$\chi_{\rm l}$ and $\log (REW)$ for our optimal values of $\Delta$$T_{\rm eff}$ and 
$\Delta \xi_{\rm t}$; immediately apparent is the fact that for the vast 
majority of lines analyzed, $\Delta$[Fe/H] is positive. We find a mean value
of $\Delta$[Fe/H] and its associated uncertainty of $+0.025\pm 0.009$ dex, where
the stated uncertainty in $\Delta$[Fe/H] was calculated via the method
described in \S 3.2, utilizing the uncertainties in $\Delta$$T_{\rm eff}$,
$\Delta \xi_{\rm t}$, and the observed scatter of $\pm 0.021$ dex in individual 
line $\Delta$[Fe/H] values. Uncertainty in $\Delta$$\log g$ produced an effect 
of $\pm 0.001$ dex in $\Delta$[Fe/H], and was not considered in the stated error 
of the mean value of $\Delta$[Fe/H]. This uncertainty is in fact dominated by 
uncertainty in $\Delta$$T_{\rm eff}$, with a $\pm 14$K change effecting a $\pm 0.008$ 
dex change in $\Delta$[Fe/H].

\section{Discussion}

\subsection{Reliability}

We note that the atmospheric parameters for 16 Cyg A and B calculated using 
the ``standard'' method stand in good agreement with previous published studies
-- a heartening fact in light of discrepancies in EW measurements\footnote
{Two typos in G98 were discovered in the course of preparing this paper: EWs
for the Fe I feature at 6710.32 \AA\ were mistakenly repeated for 6733.15 \AA\
(the latter was not measured in that study), and EWs for the Ti I feature noted
at 6126.22 \AA\ were actually from a Ti I line at 6261.11 \AA.} as well as
variations in methodology among the various authors. We believe our
absolute values are more reliable than those of other recent studies due to our
use of more Fe I lines spanning a larger range in $\chi_{\rm l}$ and EW (D00,
for example, only employed 20 Fe I lines with the smallest $\chi_{\rm l}$ and
EW values being 2.18 eV and 24 m\AA, respectively). Furthermore, using our method
of differential analysis, we are able to fine-tune the observed values of $\Delta$$T_
{\rm eff}$ and $\Delta \xi_{\rm t}$, and as a result have shown for the first time a
statistically significant difference in [Fe/H] between 16 Cyg A and B. In Figure 3 we
compare our resulting $\Delta$[Fe/H] values with those of previous studies.

The magnitude of this difference remains quite small, though, and while we have
strived to eliminate possible systematic differences in our observations and analyses
of these two stars, we must consider other potential mechanisms which might mimic true
differences in chemical abundance before we can conclude that 16 Cyg A is in fact
enhanced in Fe relative to 16 Cyg B. Most notably, measured values of [Fe/H] are known
to vary with chromospheric activity levels. A recent study of Ca II fluxes in planet-bearing
stars by Henry et al. (2000), however, shows low values of $R'_{\rm HK}$ for 16 Cyg B,
indicating little variability and surface activity. Our own spectra confirm the low
levels of Ca II H and K emission, and show no signs of significant differences between
16 Cyg A and B in this regard. Furthermore, {\it Hipparcos} data sets the photometric
variability of each of these stars at the 0.0007 magnitude level. We therefore are
confident that differences in chromospheric activity do not play a significant role
in accounting for the measured values of $\Delta$[Fe/H].

In addition, we examined possible systematic errors which might be introduced due to
assumptions in our atmospheric modeling, such as the possibility that a
difference in the temperature minimum between the two stars might yield differing
values of [Fe/H] for the same $T_{\rm eff}$ (Wallerstein, 1972). To test against this, we
compared the central absorption in the Mg I triplet near 5170 \AA, and found no
evidence supporting differences in temperature minima between 16 Cyg A and B.
We further argue that given the common ages of $9 \pm 2$ Gyr (G98) and similar values
of $T_{\rm eff}$, $\log g$, and luminosity of these two stars -- confirmed
independently from photometric analyses (c.f. F93) -- it is unlikely that any
systematic errors in this strictly differential study might be due to more subtle
model assumptions, such as that of LTE. We therefore conclude that our measured value
of $\Delta$[Fe/H] is reliable to the level of our stated uncertainty, and that the
photosphere of 16 Cyg A {\it is} significantly more iron-rich than that of 16 Cyg B.

\subsection{Possible Explanations}

An obvious explanation for the observed value of $\Delta$[Fe/H] is that it represents
a primordial difference in the chemical composition of 16 Cyg A and B. Little
high-resolution work exists on abundance differences between members of multiple star
systems, however, and models of the formation of stellar systems are currently not
sophisticated enough to address deviations in chemical abundance at the level of precision
that we report here. We intend to actively pursue the former shortcoming by employing our
differential method on a wide sample of stellar systems; however, until such primordial
fluctuations are at least empirically, if not theoretically constrained, there
is little more that can be said for or against this hypothesis.

If, on the other hand, the observed difference in [Fe/H] is not entirely primordial,
then it must be set by variations in the evolutionary history of these two stars.
A potential clue to the responsible mechanisms may lie in the aforementioned fact that
the Li abundances of 16 Cyg A and B differ by a factor of at least five (D00).
While Ryan (2000) has noted that planet-bearing stars do not as a whole appear to
differ from the field population in their values of [Li/H] -- a finding supported by the
larger sample discussed in Gonzalez et al. (2000) -- it remains a challenge for
standard theories of stellar atmospheres and evolution to explain the large difference
in [Li/H] between 16 Cyg A and B given the considerable similarity of their physical
characteristics and presumably shared environmental history.
 
In response, various authors have discussed the possibility that the chemical
composition of stellar photospheres can be affected by the presence of planetary
companions. Cochran et al. (1997) and King (1997) proposed that dynamical interactions
between a rotating star and its protoplanetary disk might significantly alter the
angular momentum evolution of the star, and hence the rate at which the star depletes
Li. This model is well supported by the recent investigations of Li and Be abundances
reported by D00, wherein these authors conclude that models of slow rotational
mixing can explain the relative abundances of Li and Be in 16 Cyg A and other
anomolously Li-rich F and G stars. Such a mechanism cannot explain the more subtle
variation in Fe which we report here, however.

Alternately, G98 proposed that the increased lithium content of 16 Cyg A's 
photosphere may be the result of that star having consumed planetary material 
in its outer convection zone. Such self-pollution by materials of either 
chondritic or gas-giant composition would indeed produce an increase in the 
Li abundance, while only slightly affecting the Fe and Be abundances
(G98, D00). If our result indicating that 16 Cyg A is slightly enhanced in Fe 
relative to 16 Cyg B is not a primordial effect, then it lends tentative support
to this scenario. From the estimates presented by G98, we calculate that the observed
difference in [Fe/H] between 16 Cyg A and B can be explained by the accretion of
2.5 M$_{\oplus}$ of chondritic or 0.3 M$_{\rm J}$ of gas giant material. The
amount of Li enhancement due to accretion of these bodies is difficult to determine,
given the marked non-linear time, $T_{\rm eff}$, and dynamical dependence
of Li depletion, along with the unkown timing of the putative accretion event(s).
Nevertheless, it remains plausible that the observed difference in Li between 16 Cyg
A and B is explainable by accretion at some intermediate age.

\acknowledgments

The authors are grateful to George Wallerstein for helpful discussions and 
support through the Kenilworth Fund of the New York Community Trust, and to 
the anonymous referee whose helpful comments greatly increased the clarity of 
our manuscript. This work has utilized the Simbad database at CDS, Strasbourg, 
France, in addition to the abstract services of ADS.

\clearpage

\begin{deluxetable}{lccccc}
\tablecaption{Fe I and II Equivalent Widths - 16 Cyg A and B}
\scriptsize
\tablewidth{0pt}
\tablehead{
\colhead{Species} & \colhead{$\lambda_{\rm o}$} & 
\colhead{$\chi_{\rm 1}$} & \colhead{$\log gf$} & \colhead{EW$_{\rm 16 Cyg A}$} & \colhead{EW$_{\rm 16 Cyg B}$}\\
\colhead{} & \colhead{(\AA)} & \colhead{(eV)} & \colhead{} & \colhead{(m\AA)} & \colhead{(m\AA)}}

\startdata
Fe I & 5044.22 & 2.85 & -2.040 & 77.2 & 77.6\nl
Fe I & 5247.06 & 0.09 & -4.930 & 70.2 & 70.7\nl
Fe I & 5322.05 & 2.28 & -2.860 & 65.8 & 65.8\nl
Fe I & 5806.73 & 4.61 & -0.900 & 60.2 & 59.2\nl
Fe I & 5852.23 & 4.55 & -1.180 & 44.3 & 44.5\nl
Fe I & 5853.16 & 1.48 & -5.280 & 9.4 & 9.6\nl
Fe I & 5855.09 & 4.61 & -1.520 & 25.2 & 25.5\nl
Fe I & 5956.71 & 0.86 & -4.550 & 55.1 & 56.5\nl
Fe I & 6024.07 & 4.55 & -0.120 & 105.1 & 109.9\nl
Fe I & 6027.06 & 4.07 & -1.090 & 69.3 & 68.0\nl
Fe I & 6034.04 & 4.31 & -2.260 & 12.9 & 13.0\nl
Fe I & 6054.08 & 4.37 & -2.200 & 11.6 & 12.6\nl
Fe I & 6056.01 & 4.73 & -0.400 & 78.3 & 78.3\nl
Fe I & 6089.57 & 5.02 & -0.860 & 40.0 & 39.2\nl
Fe I & 6093.65 & 4.61 & -1.340 & 35.3 & 35.9\nl
Fe I & 6096.67 & 3.98 & -1.810 & 41.4 & 41.9\nl
Fe I & 6098.25 & 4.56 & -1.740 & 19.4 & 19.4\nl
Fe I & 6151.62 & 2.18 & -3.290 & 55.0 & 55.3\nl
Fe I & 6157.73 & 4.07 & -1.250 & 66.7 & 66.0\nl
Fe I & 6159.38 & 4.61 & -1.870 & 14.2 & 14.2\nl
Fe I & 6165.36 & 4.14 & -1.470 & 49.0 & 48.2\nl
Fe I & 6180.21 & 2.73 & -2.610 & 58.0 & 59.7\nl
Fe I & 6200.32 & 2.61 & -2.440 & 77.2 & 75.1\nl
Fe I & 6213.44 & 2.22 & -2.660 & 87.6 & 88.2\nl
Fe I & 6226.74 & 3.88 & -2.030 & 34.1 & 34.2\nl
Fe I & 6229.23 & 2.84 & -2.820 & 42.3 & 42.3\nl
Fe I & 6240.65 & 2.22 & -3.320 & 54.2 & 55.2\nl
Fe I & 6265.14 & 2.18 & -2.570 & 90.7 & 89.7\nl
Fe I & 6270.23 & 2.86 & -2.570 & 55.4 & 56.0\nl
Fe I & 6380.75 & 4.19 & -1.320 & 57.7 & 57.5\nl
Fe I & 6385.73 & 4.73 & -1.820 & 12.6 & 11.5\nl
Fe I & 6392.54 & 2.28 & -4.010 & 20.6 & 20.5\nl
Fe I & 6498.95 & 0.96 & -4.620 & 49.7 & 51.9\nl
Fe I & 6581.22 & 1.48 & -4.660 & 27.9 & 28.2\nl
Fe I & 6591.33 & 4.59 & -1.980 & 12.7 & 11.5\nl
Fe I & 6608.04 & 2.28 & -4.010 & 19.9 & 21.2\nl
Fe I & 6627.56 & 4.55 & -1.440 & 32.6 & 32.6\nl
Fe I & 6646.97 & 2.61 & -3.850 & 12.0 & 11.3\nl
Fe I & 6653.91 & 4.15 & -2.410 & 12.0 & 12.4\nl
Fe I & 6703.58 & 2.76 & -3.010 & 42.2 & 42.8\nl
Fe I & 6710.32 & 1.48 & -4.800 & 17.2 & 19.0\nl
Fe I & 6725.36 & 4.10 & -2.180 & 20.8 & 21.3\nl
Fe I & 6726.67 & 4.61 & -1.040 & 52.3 & 52.5\nl
Fe I & 6733.15 & 4.64 & -1.450 & 30.8 & 30.7\nl
Fe I & 6739.52 & 1.56 & -4.900 & 13.0 & 13.8\nl
Fe I & 6745.98 & 4.07 & -2.770 & 9.1 & 8.7\nl
Fe I & 6746.98 & 2.61 & -4.410 & 4.7 & 5.3\nl
Fe I & 6750.16 & 2.42 & -2.620 & 79.4 & 80.4\nl
Fe I & 6752.72 & 4.64 & -1.200 & 38.9 & 40.2\nl
Fe I & 6786.86 & 4.19 & -1.950 & 31.2 & 30.5\nl
Fe I & 6820.37 & 4.64 & -1.170 & 47.9 & 47.4\nl
Fe I & 6839.84 & 2.56 & -3.360 & 36.8 & 38.2\nl
Fe I & 6855.72 & 4.61 & -1.730 & 24.1 & 23.8\nl
Fe I & 6861.95 & 2.42 & -3.800 & 23.5 & 24.3\nl
Fe I & 6862.50 & 4.56 & -1.350 & 35.0 & 35.7\nl
Fe I & 6864.32 & 4.56 & -2.300 & 7.7 & 9.0\nl
Fe I & 7498.54 & 4.14 & -2.090 & 22.4 & 22.7\nl
Fe I & 7507.27 & 4.41 & -1.050 & 64.7 & 64.5\nl
Fe I & 7583.80 & 3.02 & -1.900 & 88.9 & 89.1\nl
Fe I & 7586.03 & 4.31 & -0.180 & 117.6 & 119.8\nl
Fe II & 5234.63 & 3.22 & -2.200 & 92.5 & 89.5\nl
Fe II & 5991.38 & 3.15 & -3.480 & 38.7 & 35.0\nl
Fe II & 6149.25 & 3.89 & -2.700 & 42.4 & 40.4\nl
Fe II & 6247.56 & 3.89 & -2.300 & 60.7 & 57.0\nl
Fe II & 6369.46 & 2.89 & -4.110 & 26.2 & 23.0\nl
Fe II & 6442.95 & 5.55 & -2.380 & 6.2 & 5.3\nl
Fe II & 6446.40 & 6.22 & -1.920 & 5.3 & 4.3\nl
Fe II & 7515.84 & 3.90 & -3.360 & 17.7 & 15.8\nl
\enddata
\end{deluxetable}

\clearpage

\begin{deluxetable}{cccc}
\tablecaption{Spectroscopically Determined Physical Parameters of 16 Cyg A 
and B using ``standard'' method.}
\scriptsize
\tablewidth{0pt}
\tablehead{
\colhead{Parameter} & \colhead{16 Cyg A} & \colhead{16 Cyg B} & \colhead{Sun}}

\startdata

T$_{\rm eff}$(K) & $5745\pm40$ & $5685\pm40$ & 5777\nl
log $g$ & $4.21\pm0.07$ & $4.26\pm0.08$ & 4.44\nl
$\xi_{\rm t}$(km/s) & $0.88\pm0.06$ & $0.80\pm0.06$ & 1.00\nl
[Fe/H] & $0.10\pm0.03$ & $0.07\pm0.03$ & 0.00\nl

\enddata
\end{deluxetable}

\clearpage

\figcaption{A plot of the slopes of $\log (REW)$ (=$\log$(EW/$\lambda$)) and $\chi_{\rm
l}$ vs. $\Delta$[Fe/H] for a range of $\Delta$$T_{\rm eff}$ and $\Delta \xi_
{\rm t}$. For each series, $\log g = 4.21$ and 4.26 for 16 Cyg A and B,
respectively. Each point represents a change of 5 K in temperature difference,
from $\Delta$$T_{\rm eff} = 80$ K for the leftmost points in each series.
Shown with a cross is the (0,0) coordinate representing the formal solution;
our optimal solution is $\Delta$$T_{\rm eff} = 62$ K and $\Delta \xi_{\rm t} =
0.05$ km/s.}

\figcaption{$\Delta$[Fe/H] vs. (a) $\chi_{\rm l}$ and (b) $\log (REW)$ for our optimal 
values of $\Delta$$T_{\rm eff}$, $\Delta \xi_{\rm t}$ and $\Delta\log g$. The 
solid lines are least-squares fits to the data.}

\figcaption{A comparison of $\Delta$[Fe/H] values for 16 Cyg A and B from various published studies, with one-sigma error bars.}

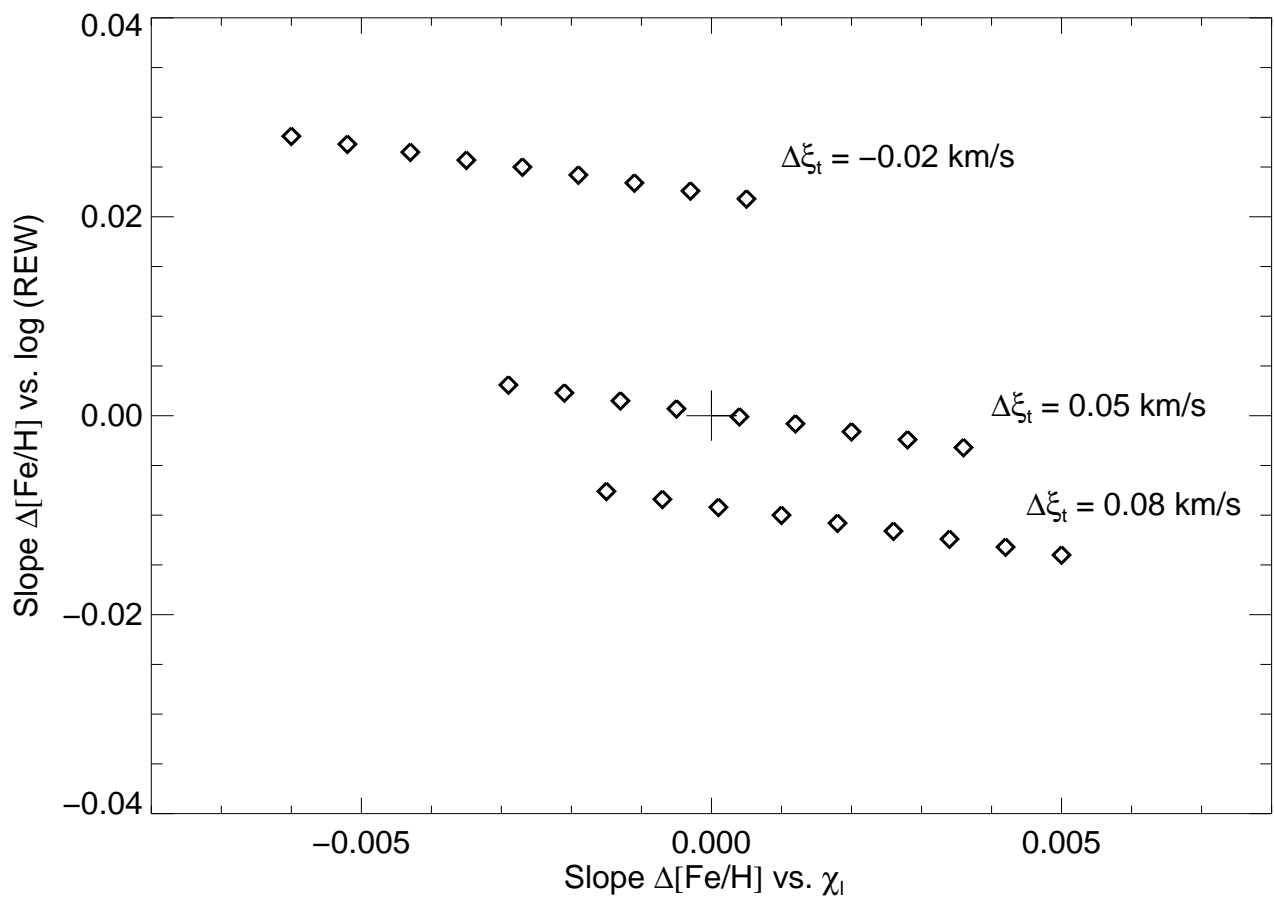
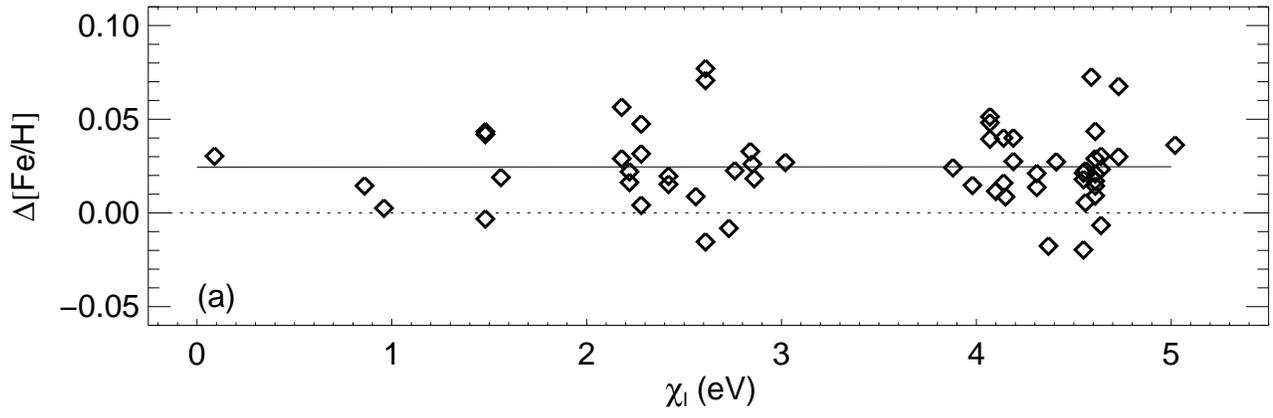
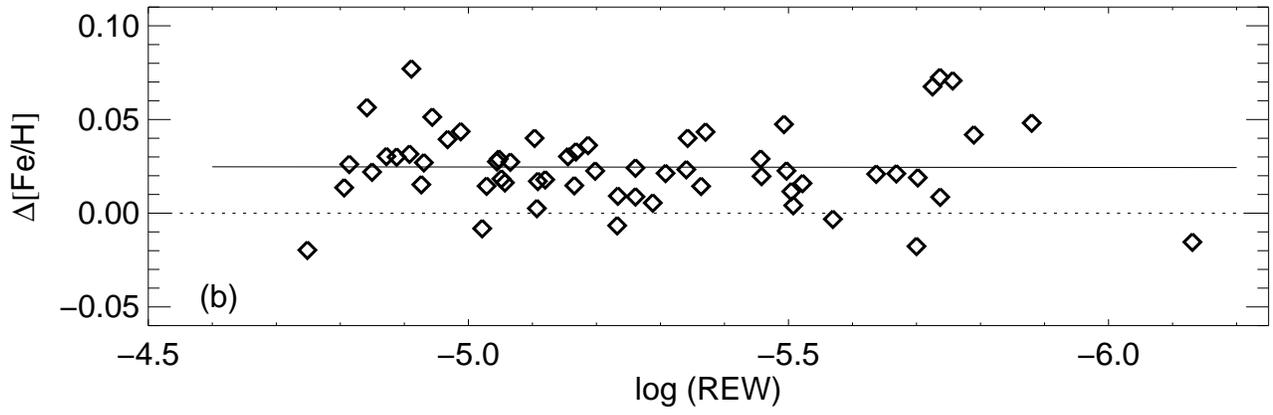
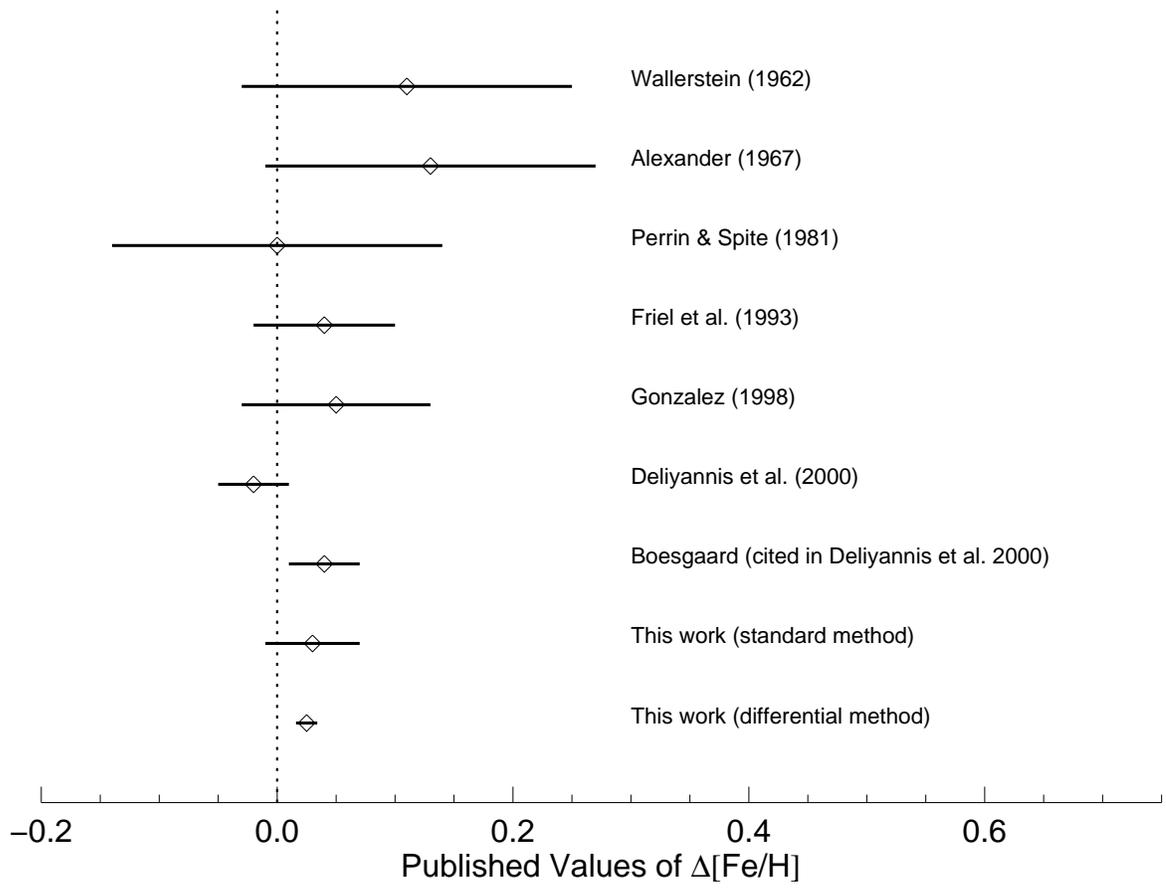
\end{document}